\documentclass[aps,prl,twocolumn,groupedaddress,showpacs]{revtex4}

\usepackage{amsmath,amssymb}
\usepackage{graphicx}
\usepackage{epstopdf}

\usepackage{amscd}
\usepackage{graphics}

\begin{document}

%\preprint{}

\title{Soliton Solutions of the KP Equation with
 V-Shape Initial Waves}


\author{Yuji Kodama}
\affiliation{Department of Mathematics, Ohio State University, Columbus, OH 43210, USA}
\author{Masayuki Oikawa}
\affiliation{Research Institute for Applied Mechanics, Kyushu University, Fukuoka 816-8580, Japan}
\author{Hidekazu Tsuji}
\affiliation{Research Institute for Applied Mechanics, Kyushu University, Fukuoka 816-8580, Japan}
%\thanks{YK }


\date{\today}

\begin{abstract}
%We consider (2,2)-soliton solutuons of the Kadomtsev-Petviashvili II %(KPII) 
%equation, that is, the solutions with two line solitons asymptotically as 
%$y \to -\infty$ and two line solitons asymptotically as $y \to \infty$. 
%It is shown that one of these solutions has all of the properties 
%which Miles' asymptotic solution for the Mach reflection of 
%a shallow water soliton has. 
We consider the initial value problems of
the Kadomtsev-Petviashvili (KP) equation for 
 symmetric V-shape initial waves
consisting of two semi-infinite line solitons with the 
same amplitude. Numerical simulations show that
the solutions of the initial value problem approach asymptotically
to certain exact solutions of the KP equation found recently in
\cite{CK08}. We then use a chord diagram to explain the asymptotic result.
We also demonstrate a real experiment of shallow water wave
which may represent the solution discussed in this Letter.

\end{abstract}

%\keywords{}

\maketitle

%\section{}
%\subsection{}
%\subsubsection{}

We consider the Kadomtsev-Petviashvili equation in
the form,
\begin{equation} 
(4u_t + 6uu_x + u_{xxx})_x + 3u_{yy} = 0,
\label{eq:KP}
\end{equation} 
which describes quasi-two dimensional, weakly nonlinear and long waves
 such as shallow water waves and ion acoustic waves \cite{KP70,OSY74,AS81}.
It is well-known that the KP equation
admits $N$-soliton solutions, each of which has asymptotically the same set of $N$ line-solitons
in both $y\to\pm\infty$.
Interaction property of two line-solitons with the same amplitude has been studied
based on a two-soliton solution which forms an ``X" shape with a phase shift due to the interaction. 
We call this two-soliton solution O-type, where ``O" stands for original. However, in 1977, Miles \cite{Miles77} pointed out that O-type soliton solution becomes singular
if the angle of the interaction is smaller than certain critical value.
Since the KP equation is supposed to give better approximations in this regime of small angles, 
it sounds strange that we do not have a reasonable solution of the equation.
Miles also found that at the critical angle the two line-solitons of the O-type solution resonantly
interact, and
a third wave (soliton) is created to make a ``Y"-shape solution, which is indeed an exact solution of the KP equation.
After the discovery of the resonant phenomena in the KP equation, several numerical and 
experimental studies were performed (see e.g. \cite{KY80, Funakoshi80, FID80, NN80}).
However, no significant progress
has been made in the study of the soliton solutions of the KP equation for almost quarter century.

In the last five years, a large variety of new soliton solutions has been found and
classified  \cite{BK03,Kodama04,BC06,CK08,CK09}. 
Those new soliton solutions enable us to describe the interaction properties of line-solitons
even in the parameter regime of small angles where O-type solution becomes singular.
In this Letter, we report how some of these new solutions appear under certain
physical settings considered in the studies on the generation of freak (or rogue) waves in shallow water 
\cite{Porubov05, OT06, TO07}.  We also present an elementary experiment
of shallow water wave demonstrating a real existence of those new solutions.

Let us first recall that $ 
u(x,y,t) = 2[\ln \tau (x,y,t)]_{xx} 
$
is a solution of  (\ref{eq:KP}), if the $\tau$-function $\tau$ is
given by  the Wronskian form,
\begin{equation} 
\tau(x,y,t)=\mbox{Wr}(f_1,\cdots,f_N), 
\label{eq:tau Wr}
\end{equation}
where the
functions $f_i$'s satisfy
the {\it linear} equations, $
f_y = f_{xx}, ~ f_t = -f_{xxx} 
$ (see e.g. \cite{H04}). 
%A simple solution of (\ref{eq:system for f}) is  
%\begin{equation} 
%$f = e^{\theta}, \ \ \theta = k x + k^2 y - k^3 t + \theta_0
%\label{eq:theta}
%\end{equation} 
%with constants $k$ and $\theta_0$. 
The soliton solutions we consider are generated by
\begin{equation} 
f_i = \sum_{j=1}^{M} a_{ij}e^{\theta_j}, \quad
\theta_j = k_j x + k_j^2 y - k_j^3 t.
\label{eq:fis}
\end{equation}    
Here the coefficient matrix $A=(a_{ij})$ is a constant $N\times M$ matrix,
and $k_j$ are constants with the ordering $k_1 < 
\cdots < k_M$. Thus each solution $u(x,y,t)$ is completely determined by
the $A$-matrix and the $k$-parameters.

\begin{figure}[t]
%\vspace{-0.5cm}
\begin{center}
%\hspace{-1.2cm}
 \includegraphics[width=38.0mm]{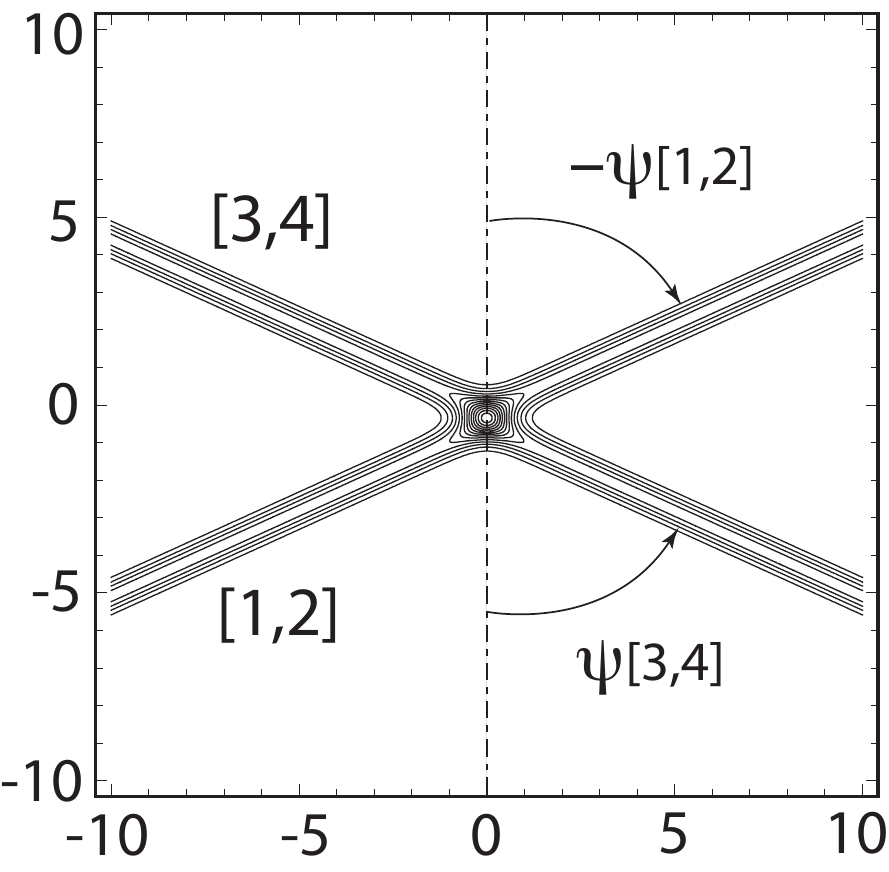}\quad\quad
 \includegraphics[width=38.0mm]{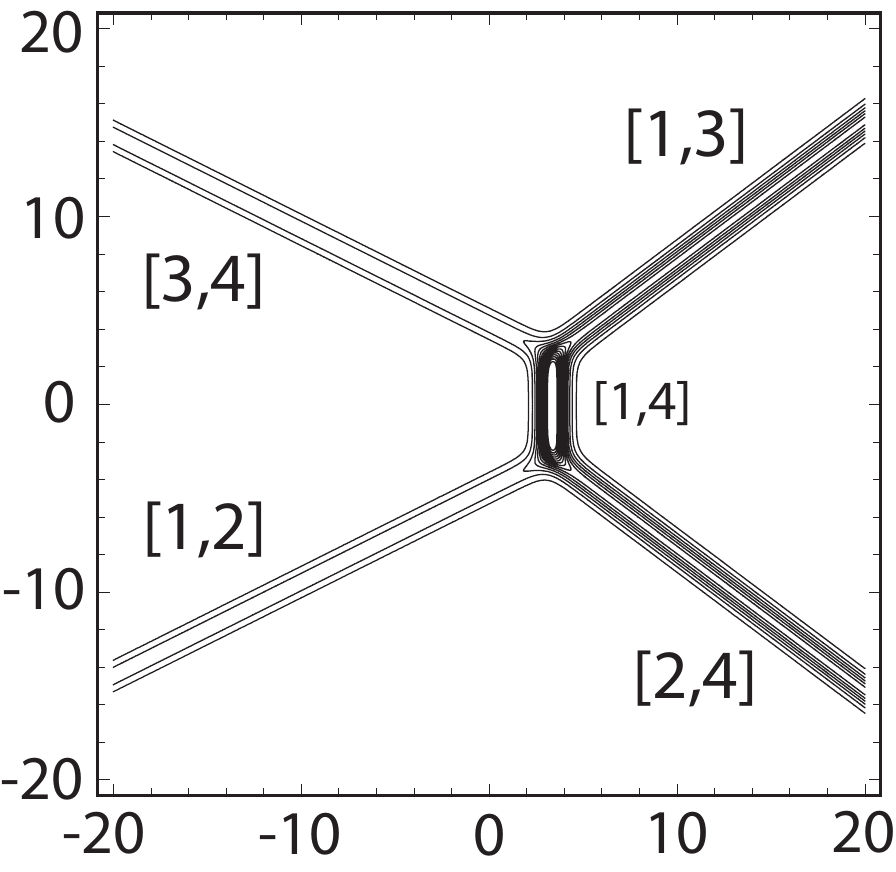}\\
 \medskip
~~ \includegraphics[width=40.0mm]{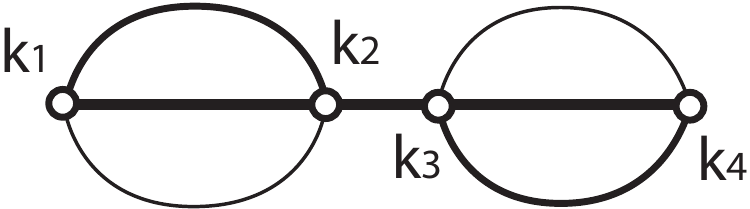}\quad\quad
\raisebox{-2.0mm}{ \includegraphics[width=35.0mm]{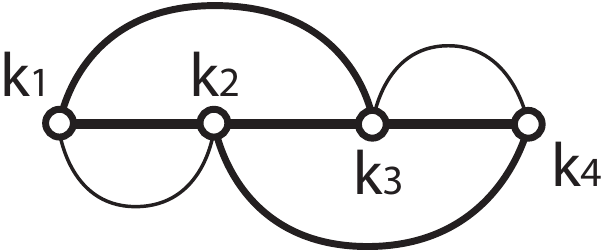}}
 \caption{Contour plots and the chord diagrams of soliton solutions. Left: O-type 
 solution. Right: (3142)-type solution. Each $[i,j]$ 
denotes the $[i,j]$-soliton. The length of $[1,4]$-soliton changes in $t$.
The upper (lower) chords represent the solitons in $y>0~(y<0)$.
The thicker chords correspond to the solitons in the right side ($x>0$).}
 \label{fig:Fig1}
 \end{center}
 \vspace{-5mm}
\end{figure}

One soliton solution is then obtained in the case with $N=1$ and $M=2$:
We have $\tau = e^{\theta_i} + e^{\theta_j} $  for $k_i<k_j$, and $u=2(\ln\tau)_{xx}$ gives
\begin{equation} 
u = \frac{1}{2}(k_j-k_i)^2\mbox{sech}^2\frac{1}{2}(\theta_j-\theta_i).
\label{eq:1-soliton-sol}
\end{equation}
The line $\theta_i=\theta_j$ represents the ridge of the soliton, 
and we refer to this {\it line}-soliton as $[i,j]$-soliton. 
The amplitude $\alpha[i,j]$ and the inclination $\gamma[i,j]:=\tan\psi[i,j]$ of this 
soliton are given by 
\begin{equation} 
\alpha[i,j] = \frac{1}{2}(k_j-k_i)^2, \quad  
\gamma[i,j] = k_i + k_j,   
\label{eq:1-soliton-amp-dir}
\end{equation} 
where $-\pi/2 < \psi[i,j] < \pi/2$ is the angle of the line soliton
 measured counterclockwise from the $y$-axis. This angle also represents 
the propagation direction of the line soliton (see Fig.\ref{fig:Fig1}).

In the case with $N=2, M=4$, using the Binet-Cauchy theorem for the determinant,
the $\tau$-function  $\tau={\rm Wr}(f_1,f_2)$ 
can be expressed in the form,
\begin{equation}
\tau = \sum_{1 \le i<j \le 4}\xi(i,j) E(i,j),
\label{eq:tau-function3}
\end{equation}
where $\xi(i,j)$ is the $2 \times 2$ minor consisting of   
$i$-th and $j$-th columns of the matrix $A$, and $E(i,j)=(k_j-k_i)e^{\theta_i+\theta_j}$. 
For the regular solutions, we require all of these minors 
to be non-negative (note $E(i,j)>0$).

As was shown in \cite{CK08,BC06,CK09}, each $\tau$-function (\ref{eq:tau-function3}) 
generates a soliton solution which consists of at most
two line-solitons for both $y \to \pm\infty$.
In the cases where the $\tau$-function (\ref{eq:tau-function3}) has at least 4 terms
(i.e. $A$ is {\it irreducible}, see \cite{BC06,CK08}), we have seven different types of
soliton solutions.
Two of them are usual two-soliton solutions, O-type 
and P-type (this type fits better in the {\it physical} setting
for the KP equation), and they are steady translational solutions with an ``X"-shape.
The other five are non-stationary solutions.

 In particular, we consider the following two types which are relevant to the solutions
 of the initial value problems considered in this Letter:
 One is the O-type solution which consists of  two line-solitons of $[1,2]$ and $[3,4]$
for $y \to \pm\infty$. Other one is non-statinary, and it consists of
$[1,3]$ and $[3,4]$ line-solitons for $y \to +\infty$ and 
$[1,2]$ and $[2,4]$ line-solitons for $y \to -\infty$. 
Let us call this soliton $(3142)$-type, since those four line-solitons
represent a permutation $\pi=\binom{1~2~3~4}{3~1~4~2}$.
Figure \ref{fig:Fig1} illustrates the contour plots of
 O-type and (3142)-type 
solutions in the $xy$-plane, and the corresponding
chord diagrams representing each soliton as a chord
joining a pair of $k_i$'s. Note in particular that the length and the location of each chord give the amplitude and
the angle of the corresponding soliton (cf. (\ref{eq:1-soliton-amp-dir})).
The $A$-matrices for those solutions are respectively given by 
\begin{equation}
A_{\rm O}=
\begin{pmatrix}
1 & a & 0 & 0 \cr
0 & 0 & 1 & b\cr
\end{pmatrix}
, \ \ A_{(3142)}=
\begin{pmatrix}
1 & a & 0 & -c \cr
0 & 0 & 1 & b\cr
\end{pmatrix}
,  
\label{eq:O-3142matrix}
\end{equation} 
where $a, b, c > 0$ are constants determining the locations of the solitons (see \cite{CK09}). Notice that the $\tau$-function  for the (3142)-type contains 5 exponential terms,
\begin{align}
\tau \!&=\!(k_3\!-\!k_1)e^{\theta_1+\theta_3} 
 \!+\! (k_4\!-\!k_1)be^{\theta_1+\theta_4} 
 \!+\! (k_3\!-\!k_2)ae^{\theta_2+\theta_3}\nonumber \\ 
 \!&+\! (k_4\!-\!k_2)abe^{\theta_2+\theta_4} \!+\! (k_4\!-\!k_3)
ce^{\theta_3+\theta_4},
\label{eq:3142-tau}
\end{align}
and the $\tau$-function for O-type with $c=0$ in (\ref{eq:3142-tau}) contains only 4 terms.

Let us fix the amplitudes and the angles of the solitons in the positive $x$ regions
for both O- and (3142)-types, so that those solutions are symmetric with respect to
the $x$-axis (see Fig.\ref{fig:Fig1}):
\begin{align}
\alpha&\equiv\left\{\begin{array}{l}\alpha[1,2]=\alpha[3,4] \quad {\rm (for~O-type)}\\
\alpha[1,3]=\alpha[2,4]\quad {\rm( for~(3142)-type)}\end{array}\right.\\
\gamma&\equiv\left\{\begin{array}{l}-\gamma[1,2]=\gamma[3,4]\quad {\rm (for~O-type)}\\
-\gamma[1,3]=\gamma[2,4]\quad {\rm (for~(3142)-type)}\end{array}\right.
\end{align}
Then from (\ref{eq:1-soliton-amp-dir}),
one can find the $k$-parameters  in terms of $\alpha$ and $\gamma$
with $k_1=-k_4$ and $k_2=-k_3$ (due to the symmetry):
In the case of O-type, we have 
\begin{equation}
k_1=-\gamma/2-\sqrt{\alpha/2}, \quad 
k_2=-\gamma/2+\sqrt{\alpha/2}.
\label{eq:values of k (O)}
\end{equation}  
The ordering 
$k_2<k_3$ then implies $\gamma > \sqrt{2\alpha}$.

On the other hand, for the (3142)-type, we have    
\begin{equation}
k_1=-\gamma/2-\sqrt{\alpha/2}, \quad 
k_2=\gamma/2-\sqrt{\alpha/2}.
\label{eq:values of k}
\end{equation} 
The ordering $k_2<k_3$  implies
$\gamma < \sqrt{2\alpha}$.

Thus, if all the solitons in the positive $x$ -region have the same amplitude $\alpha$
for both O- and (3142)-types, then an O-type solution arises when $\gamma>\sqrt{2\alpha}$,
and a (3142)-type when $\gamma<\sqrt{2\alpha}$. 
Then the limiting value at $k_2=k_3$ $(=0)$
 defines the critical angle, i.e.
\begin{equation}\label{critical}
\gamma_c:=\sqrt{2\alpha}.
\end{equation}
Note from (\ref{eq:3142-tau}) that at the critical angle, i.e. $k_2=k_3$, the $\tau$-function
has only three exponential terms, and this gives a ``Y"-shape resonant solution
as Miles noted \cite{Miles77}.

One can also see from (\ref{eq:1-soliton-amp-dir}) that for (3142)-type solution, the solitons
in the negative $x$-region are smaller than those in the positive region, i.e.
$\alpha[3,4]=\alpha[1,2]=\gamma^2/2 < \alpha$, and the angles of those in the negative $x$-regions
do not depend on $\gamma$ and
$\gamma[3,4]=-\gamma[1,2]=\gamma_c$. 
Two sets of three solitons $\{[1,3],[1,4],[3,4]\}$ 
and $\{[2,4],[1,4],[1,2]\}$ are both in the soliton 
resonant state, that is, they are resonant triplets.
These properties of the (3142)-type solution 
are the same as those of Miles' asymptotic solution 
for the Mach reflection of a shallow water 
soliton\cite{Miles77}.

We now discuss the numerical simulations of the KP equation with
 V-shape initial waves with $\alpha=2$ and one free parameter $\gamma$ ($0<\gamma<\pi/2$),
\begin{equation}
u(x,y,0) =
2\,\mbox{sech}^2(x - \gamma |y|) .
\label{eq:initial}
\end{equation}
This initial wave form has been considered in the study on an oblique reflection of a solitary wave 
in two layer fluid due to a rigid wall along 
the $x$-axis\cite{TO01,TO07}. 
The numerical simulations are based on a spectral method with window-technique
where the boundary of the computation domain is patched by
the corresponding one-soliton solutions (the details will be published elsewhere).

\begin{figure}[t]
\begin{center}
\includegraphics[width=75.0mm]{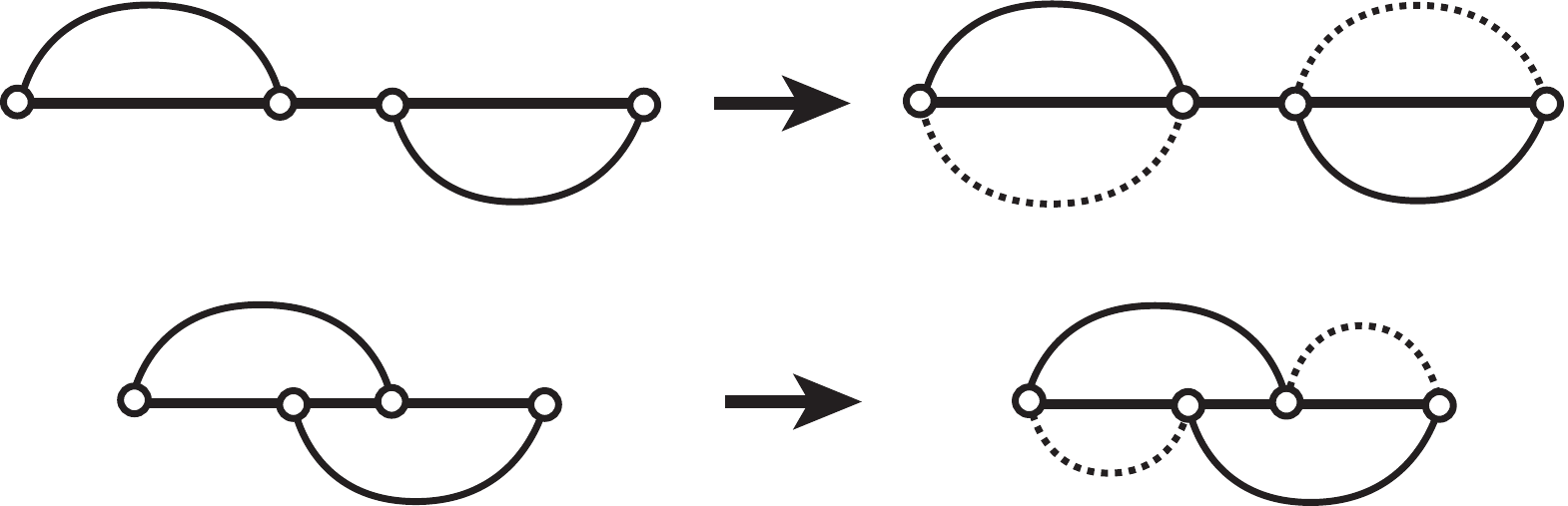}%
\end{center}
\vspace{-5mm}
\caption{Completion of the chord diagrams. The left diagrams show the initial V-shape waves with
the same amplitude for O-type (top) and (3142)-type (bottom).
The right diagrams represent the asymptotic solutions.}
\label{fig:CE}
\vspace{-2mm}
\end{figure}

In terms of the chord diagrams, the V-shape initial waves are represented by 
the partial chords in Fig.\ref{fig:Fig1} with the thicker chords.
Then the main goal in this Letter is to show that the asymptotic solutions of the simulations
are given by O-type solution for $\gamma>\gamma_c$ and (3142)-type
for $\gamma<\gamma_c$. Namely, the partial chords
are asymptotically getting to be the corresponding complete chords (see Fig.\ref{fig:CE},
and the discussion in the end of this Letter). 

\begin{figure}[t]
\begin{center}
\includegraphics[width=82.0mm]{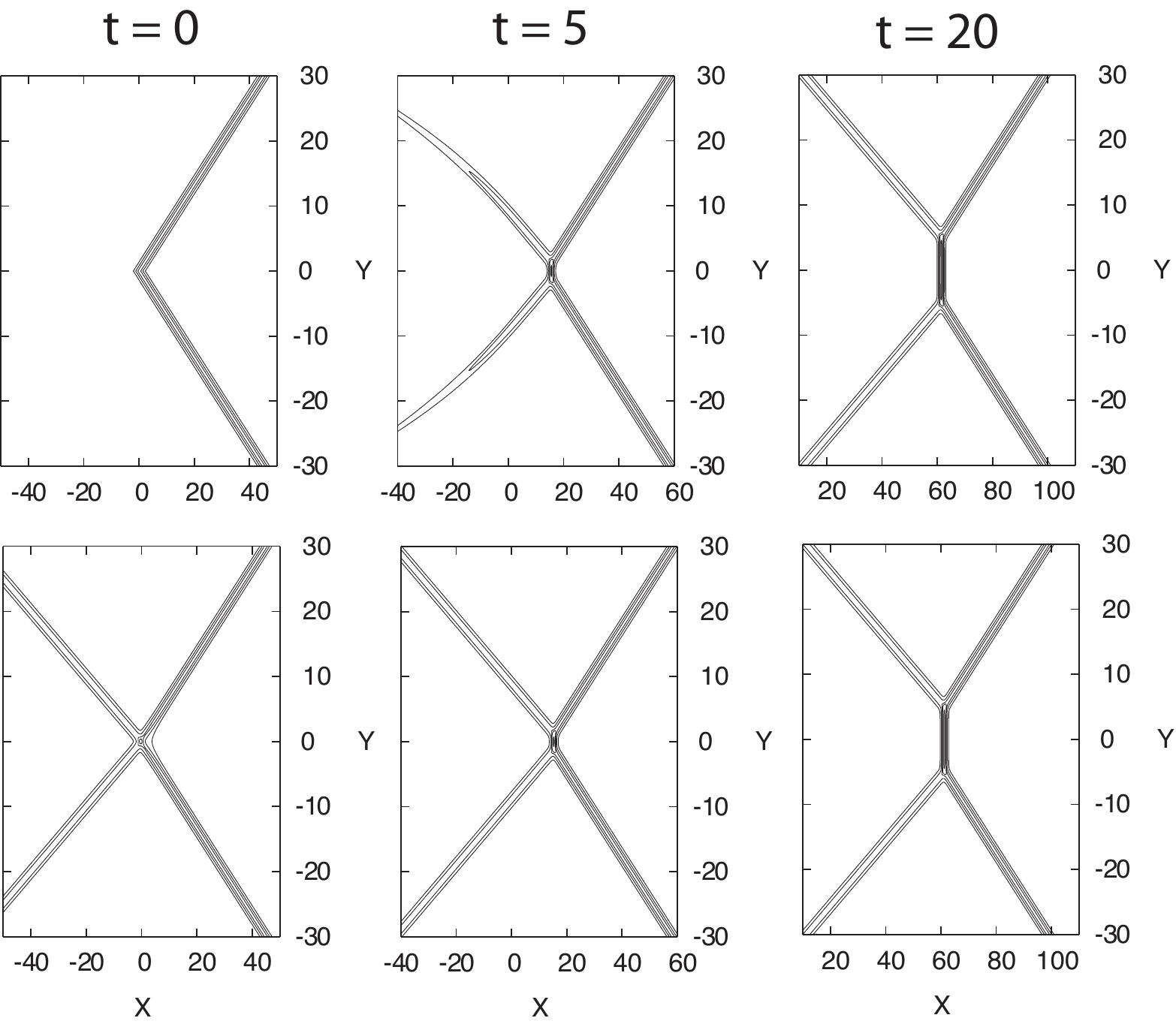}%
\end{center}
\vspace{-5mm}
\caption{The contour plots of the solutions. Upper: Numerical solution for V-shape initial 
wave with $\alpha=2$ and $\gamma=1.5$.
Lower: (3142)-type solution at the same times, and the parameters are
 $(k_1,\ldots,k_4)=
(-7/4,-1/4, 1/4, 7/4)$ and $(a,b,c)=(4, 4/7, 4/3)$ in the $A$-matrix,
so that all solitons meet at the origin at $t=0$ (see \cite{CK09}).}
\label{fig:Fig2}
\vspace{-3mm}
\end{figure}
Figure \ref{fig:Fig2} shows the contour plots of the 
numerical solutions for $\gamma=1.5 ~(< \gamma_c)$
and the (3142)-type solution with the same $\gamma$.  Note that 
the waves in the trailing part of the solution
are fully developed at $t=20$ within the sight of numerical domain.   
At $t=20$ the numerical solution is in remarkable agreement with 
the $(3142)$-type solution:
%%%%    
\begin{figure}[t]
\begin{center}
\includegraphics[width=90.0mm]{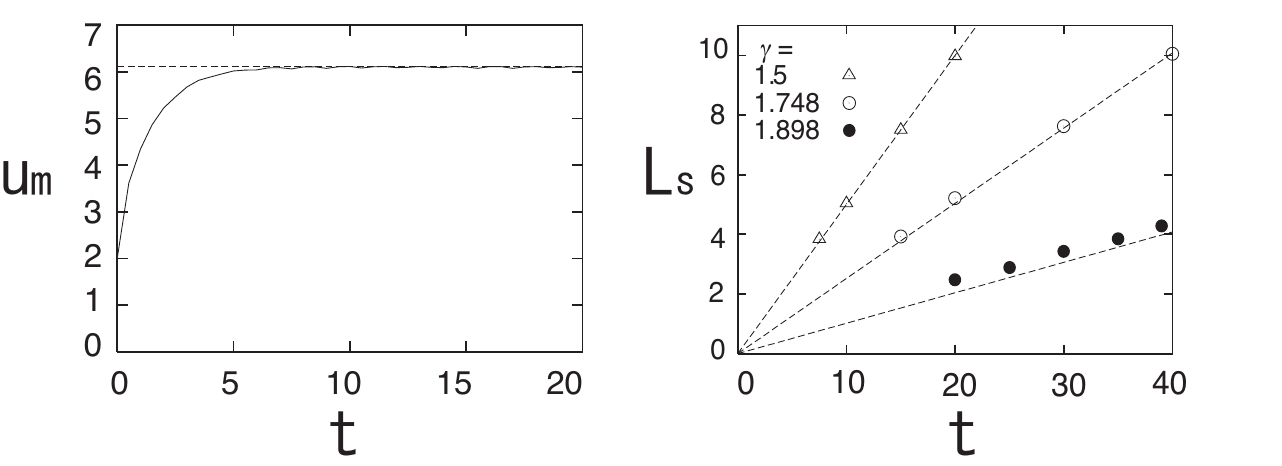}%
\end{center}
\vspace{-5mm}
\caption{ 
Left: The time evolution of the stem amplitude $u_m$ (solid line) for 
$\alpha = 2, \ \gamma = 1.5$. The dashed line denotes $\alpha[1,4]=6.125$.
Right: The time evolution of the stem length $L_s$ for $\alpha = 2$ and 
$\gamma=1.5, 1.748, 1.898<\gamma_c=2$. The dashed lines denote (\ref{eq:stemL}).}
\label{fig:Fig3}
\vspace{-5mm}
\end{figure}  
%%%%        
%%%%
The waves in the trailing part in the negative $x$-region are almost
identical with $[3,4]$- and $[1,2]$-solitons in 
$y>0$ and $y<0$ respectively. From the corner of the V-shape, 
a new wave is generated. This wave is called {\it stem}, and can be identified as 
the $[1,4]$-soliton (see Fig.\ref{fig:Fig1}).     
The time evolution of the stem amplitude $u_m$ is shown 
in Fig.\ref{fig:Fig3}, and it approaches to the asymptotic value,
$\alpha[1,4] = (\gamma+\sqrt{2\alpha})^2/2=6.125$. 
The time evolutions of the stem length $L_s$ are also shown 
in Fig.\ref{fig:Fig3}. Identifying the stem as [1,4]-soliton,
the formula of $L_s$ can be derived. The ridges of 
$[1,3]$ and $[1,4]$ solitons are given by $\theta_1=\theta_3$ and $\theta_1=\theta_4$,
which lead to
\begin{equation}
x-\gamma y =\frac{1}{4}(\gamma_c^2+3\gamma^2)t, \quad
x=\frac{1}{4}(\gamma_c+\gamma)^2 t .
\label{eq:ridges}
\end{equation}
Then the stem length $L_s$ is given by $L_s=2y$ with $y$ at the intersection
point of those lines (\ref{eq:ridges}), i.e.
\begin{equation}\label{eq:stemL}
L_s=(\gamma_c-\gamma)t.
\end{equation}
We performed simulations for the cases with $\gamma=1.5, ~1.748, ~1.898<\gamma_c$,
and found that  the solutions 
agree generally well with the corresponding (3142)-type 
solutions. However, as seen in Fig.\ref{fig:Fig3}, 
the stem length $L_s$ of the numerical solution is slightly longer than 
the expected value (\ref{eq:stemL}) when $\gamma$ approaches to $\gamma_c$.

For $\gamma > \gamma_c=2$, we performed the simulations for the cases with 
$\gamma=2.1, 2.207, 2.367, 2.5$, and confirmed that the solutions approach to 
O-type solutions with the same $\gamma$. Namely the waves in the trailing part in
$x<0$ are identified as $[1,2]$- and $[3,4]$-solitons with a phase shift $\delta_x$
 which is given by \cite{Miles77}
\begin{equation}
\delta_x = \gamma_c^{-1}\ln[\gamma^2/(\gamma^2-\gamma_c^2)]. 
\label{eq:shift}
\end{equation}  
  Also using the exact solutions of O-type and (3142)-type, one can obtain the formula $u_a$
  in terms of $\gamma$
%%%%    
\begin{figure}[t]
\begin{center}
\includegraphics[width=85.0mm]{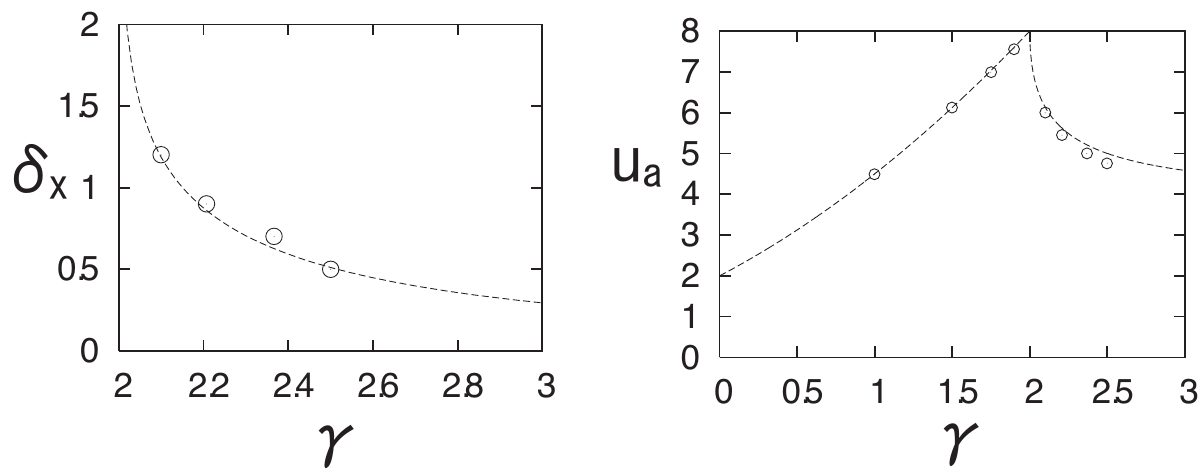}%
\end{center}
\vspace{-5mm}
\caption{ 
Left: The phase shift $\delta_x$ for O-type solution. The dashed line denotes (\ref{eq:shift}). 
The circle denotes numerical results for $\gamma = 2.1, 2.207, 2.367, 2.5$. 
Right: The asymptotic value $u_{\rm a}$ 
of the maximum of $u(x,y,t)$. The dashed line denotes (\ref{eq:Max-amp}).
 The circle denotes numerical results. 
% for $\gamma= 1, 1.5, 1.748, 1.898, 
%2.1, 2.207, 2.367, 2.5$.  
}
\label{fig:Fig5}
%\vspace{-5mm}
\end{figure}  
%%%% 
 \cite{Miles77},
\begin{equation}\label{eq:Max-amp}
u_{\rm a}=\left\{\begin{matrix}(\gamma +2)^2/2 &{\rm  for} &\gamma < \gamma_c=2,\\[0.5ex]
8/(1+e^{-\delta_x})&{\rm for} &\gamma > \gamma_c=2.
\end{matrix}\right.
\end{equation}
Figure \ref{fig:Fig5} illustrates the numerical results for the phase shift and the asymptotic maximum amplitude $u_a$ with the formulae (\ref{eq:shift}) and (\ref{eq:Max-amp}).

At the critical angle $\gamma=\gamma_c=2$, the maximum value of $u$
is given by $u_m=4\alpha=8$ (cf. (\ref{eq:Max-amp})).
%%%%    
\begin{figure}[t]
\begin{center}
%\smallskip
\includegraphics[width=85.0mm]{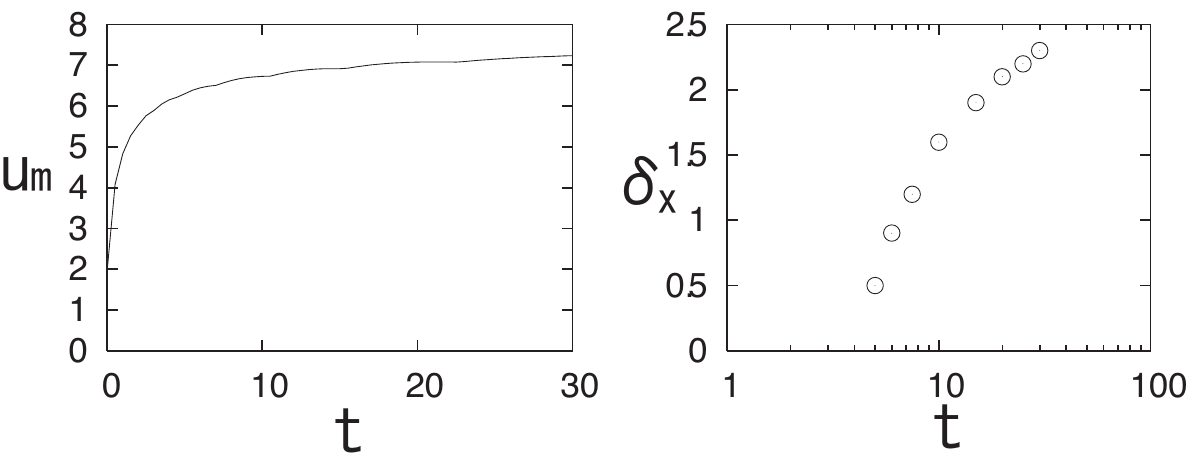}%
\end{center}
\vspace{-5mm}
\caption{ 
Left: The time evolution of the maximum $u_m$ of $u(x,y,t)$ at
the critical angle $\gamma=\gamma_c=2$. 
Right: The time evolution of the phase shift $\delta_x$ 
at the critical angle $\gamma=2$. }
\label{fig:Fig6}
%\vspace{-5mm}
\end{figure}  
%%%%
The numerical simulation shows that the amplitude approaches very 
slowly to the asymptotic value 8 (Fig.\ref{fig:Fig6}). 
In the limit $\gamma\to 2^+$ of O-type solution, the phase shifts become infinity. 
On the other hand, in the limit $\gamma\to 2^-$ of the (3142)-type solution, 
the stem length $L_s$ approaches to zero.  
We could not find an exact description of the solution at the critical angle, but      
the simulation indicates that the phase shift (or the stem length)
seems to have a logarithmic increase (Fig.\ref{fig:Fig6}).

\begin{figure}[t!]
\begin{center}
%\smallskip
\quad\includegraphics[width=35.0mm]{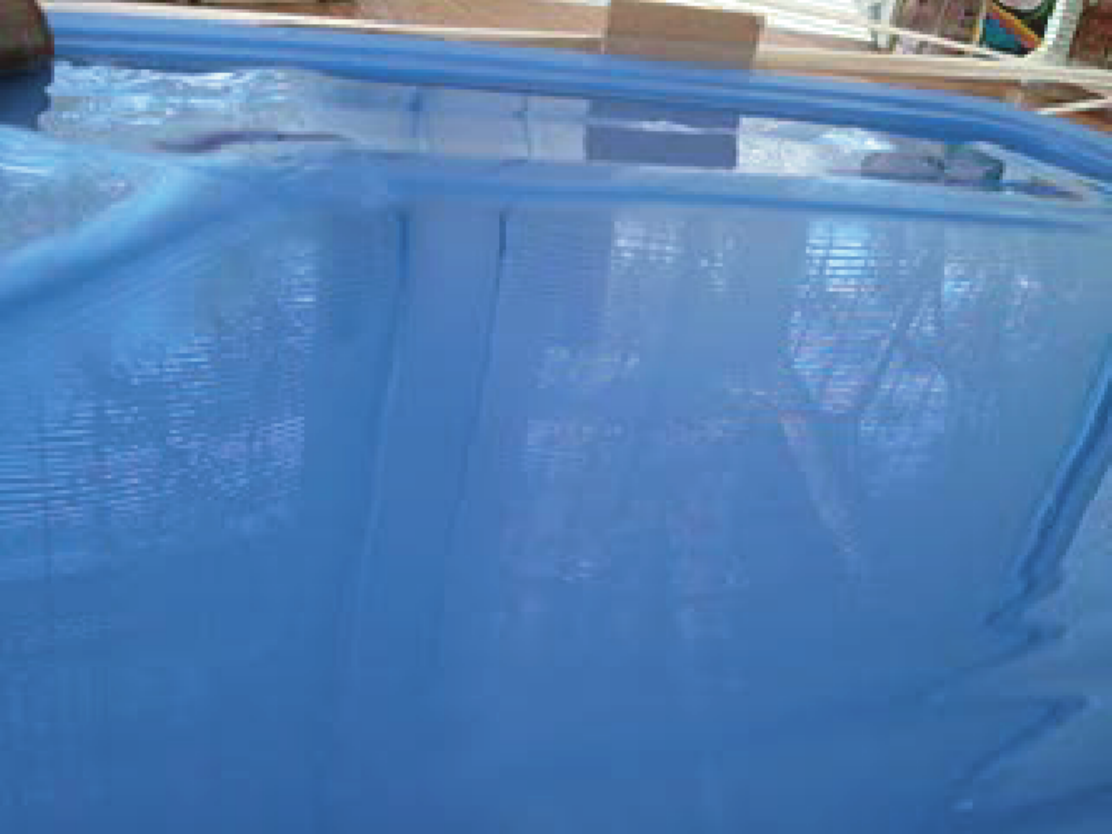}\quad
\raisebox{-2mm}{\includegraphics[width=43mm]{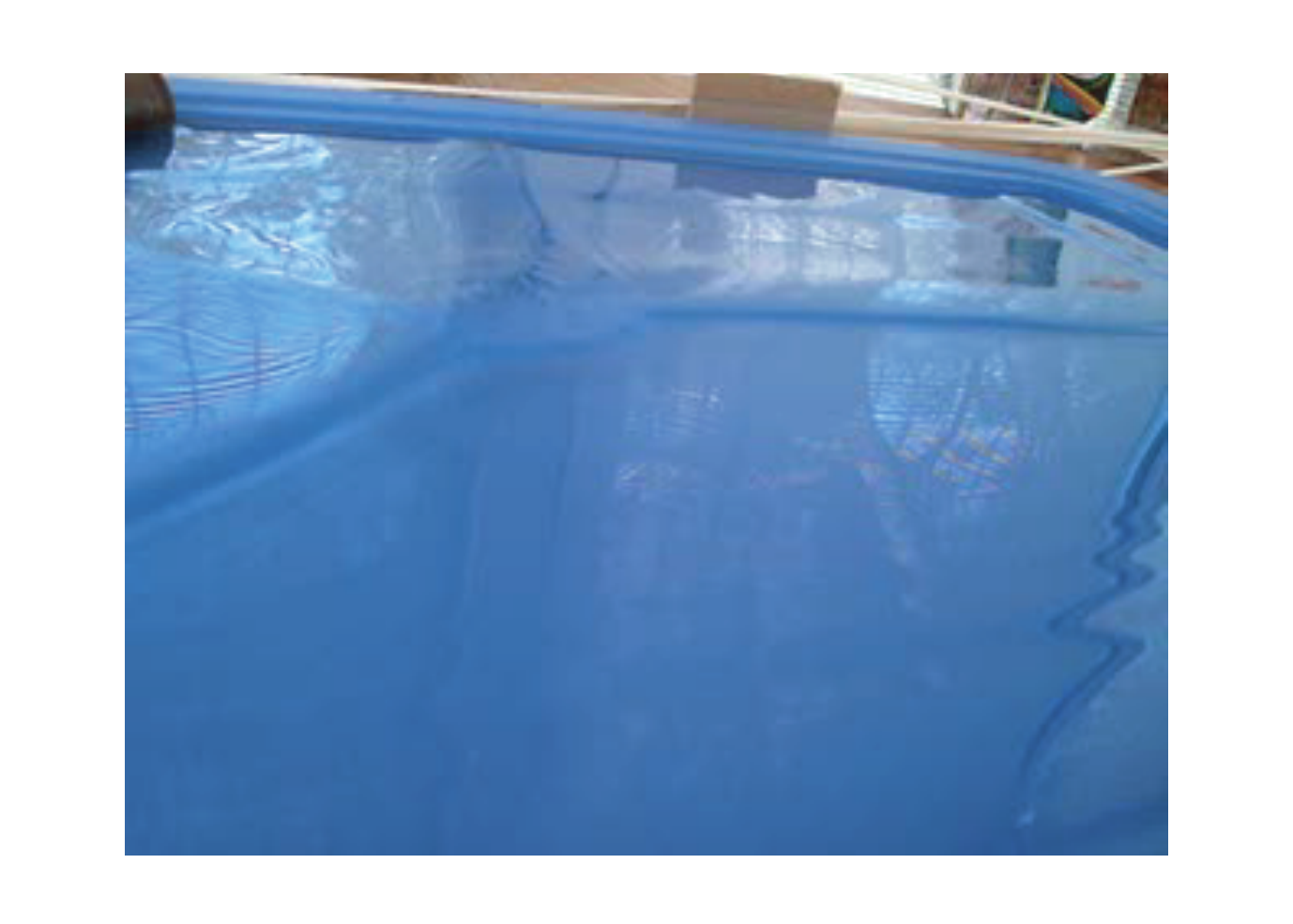}}
\end{center}
\vspace{-5mm}
\caption{A real experiment  showing a (3142)-type solution.}
\label{fig:Fig7}
\vspace{-5mm}
\end{figure}

Let us now make a summary and some discussion:   We performed several numerical simulations
 of the KP equation with symmetric V-shape initial waves.
Then we found that the solutions
are asymptotically approaching to either (3142)-type solutions or O-type solutions 
depending on the angles of the V-shape.
Those solutions are expressed by the chord diagrams, equivalently
the permutations (see Fig.\ref{fig:Fig1}).  As shown in \cite{CK09}, 
the soliton solutions generated by the $\tau$-functions in the form (\ref{eq:tau Wr})
are uniquely expressed by the permutations.
Then in order to explain the asymptotic solutions, we first express each initial V-shape wave
as a (sub) chord diagram consisting of two chords which correspond to the semi-infinite solitons
of the V-shape. Then we observed that the asymptotic solution is given by
a chord diagram which is the {\it completion} of the sub chord diagram.
Here the completion means that the resulting chord diagram should give the unique permutation
and the corresponding solution has the {\it smallest} total length of the chords.
For example, for the (sub) chord diagram of V-shape with $\gamma>\gamma_c$, there are {\it two} complete 
chord diagrams having this sub diagram (they are of O-type and (2413)-type).
The total length of solitons in (2413)-type is larger than the O-type one. Thus the completion
of the sub chord diagram is the diagram of the O-type.
This completion can be also applied for more general cases of V-shape initial waves with
different amplitudes. The details will be reported elsewhere. 
Finally, in Fig.\ref{fig:Fig7}, we present an elementary ({\it desktop}) experiment of a shallow water wave. The size of the tank is 
37cm$\times $65cm with 2cm depth. 
The waves are generated by shifting the tank in a diagonal direction. The resulting wave is similar to
(3142)-type. The experiment is easy, and of course, we can also get ``Y"-shape and O-type as well.

\medskip
\noindent
{\bf Acknowledgement}: YK is partially supported by NSF grant DMS0806219.

\end{document}